\documentclass[preprint]{revtex4}
\usepackage[T1]{fontenc}
\usepackage[dvips]{graphicx}
\usepackage{rotating}
\usepackage{bm}        
\usepackage{amssymb}   
\usepackage{amsmath} 
\usepackage{bm}
\def\etal{{\it et al.}}

\def\jcp#1#2#3{J.~Chem.~Phys.~{\bf #1},\ #2\ (#3)}

\def\pra#1#2#3{Phys.~Rev.~A~{\bf #1},\ #2\ (#3)}
\def\prl#1#2#3{Phys.~Rev.~Lett.~{\bf #1},\ #2\ (#3)}

\def\k1{k_1}
\def\k2{k_2}
\def\q1{q_1}
\def\q2{q_2}

\def\({\left (}
\def\){\right )}
\def\[{\left [}
\def\]{\right ]}

\newcommand{\beq}{\begin{equation}}
\newcommand{\eeq}{\end{equation}}

\begin{document}
\date{\today}
\flushbottom \draft
\title{
Controlling electronic spin relaxation of cold molecules with electric fields
}
\author{ T. V. Tscherbul and R. V. Krems
}
\affiliation{
Department of Chemistry, University of British Columbia, Vancouver, B.C. V6T 1Z1, Canada
}
\begin{abstract}
We present a theoretical study of atom - molecule collisions in superimposed electric and magnetic fields and show that dynamics of electronic spin relaxation in molecules at temperatures below 0.5 K can be manipulated by varying the strength and the relative orientation of the applied fields. The mechanism of electric field control of Zeeman transitions is based on an intricate interplay between intramolecular spin-rotation couplings and molecule-field interactions. We suggest that electric fields may affect chemical reactions through inducing nonadiabatic spin transitions and facilitate evaporative cooling of molecules in a magnetic trap.

\end{abstract}

\maketitle

\clearpage
\newpage

Building a quantum computer \cite{zoller}, establishing  the 
time-reversal symmetry of nature \cite{hinds}, and achieving external control over chemical reactions and molecular dynamics processes \cite{moshe}
are some of the most important fundamental problems of contemporary physics and chemistry. 
It has been recently realized that efficient quantum information processing \cite{demille,mitia,zoller2},  
precision symmetry measurements  \cite{hinds,demille2,john,demille3} and external field control 
of molecular collisions \cite{irpc} may all become possible using molecular ensembles cooled to temperatures below 1 Kelvin. 
In particular, it was suggested that molecules with non-zero electronic spin trapped on an 
optical lattice may act as qubits of a quantum computer \cite{mitia,zoller2} or used for 
measurements of the electric dipole moment (EDM) of the electron \cite{hinds,john}, which
may provide a test of the time-reversal symmetry. 
Both the quantum computation schemes and the precision measurements rely on a coherent superposition of molecular spin states prepared or exploited in the presence of magnetic and electric fields. Molecular collisions destroy the coherence by inducing spin relaxation. Electromagnetic fields modify the structure of molecules and may thus affect collisional spin relaxation. In this work, we analyze the effect of electric fields on collisional spin relaxation in cold molecules. Our results lead us to propose several new mechanisms for controlling molecular collisions and, possibly, chemical reactions with superimposed electric and magnetic fields.

Collisions of molecules in external electric and magnetic fields have been studied by Bohn and coworkers \cite{bohn_e,bohn_m} and by our group 
\cite{irpc,krems,krems2}. These studies demonstrated that dynamics of molecules at zero absolute temperature may be
sensitive to the magnitude of an applied field.
All of these papers, however, focussed on molecular collisions at yet unrealistic temperatures of less than 1 mK and the effects observed may not be present in warmer gases. Here, we explore the effect of combined electric and magnetic fields on collisions of cold ($\sim$0.5 K) molecules.  
These temperatures are relevant for the EDM measurement experiments \cite{john} and can be easily achieved with cryogenic cooling techniques \cite{buffer-gas-loading}. Expanding on the work of Friedrich and Herschbach \cite{bretislav}, we examine the interplay of electric and magnetic fields and the intramolecular spin-rotation interactions as they determine spin relaxation in inelastic collisions. In particular, we explore the effect of the relative orientation of magnetic and electric fields on collisions and propose that rotating electric fields may alter the dynamics of spin transitions. This work is the first study of collisions in crossed electric and magnetic fields. The symmetry of the collision problem is completely destroyed if the magnetic and electric fields are rotated and we discuss the corresponding complications arising in the quantum scattering theory.

We consider collisions of CaH and CaD molecules in the electronic ground state $^2\Sigma$ with $^3$He atoms.  Magnetic fields split the rotational ground state ($N=0$) of a $^2\Sigma$ molecule into two Zeeman energy levels, corresponding to the spin-up and spin-down orientations of the electronic spin \cite{bretislav, roman2}.
We solve the time independent scattering problem in the fully uncoupled representation of the wavefunction \cite{krems2}. The diatomic molecule is described by the Hamiltonian 
\beq\label{e2}
\hat{H} =
\hat{H}_{\rm rv}+ \gamma {\hat S}\cdot{\hat N} - {\hat E}\cdot{\hat d} + 2\mu_{\text{B}}{\hat B}\cdot{\hat S},
\eeq
where $\hat{H}_{\rm rv}$ determines the ro-vibrational structure of the field-free molecule \cite{mizushima}, $\gamma$ is the constant of the spin-rotation interaction between the rotational angular momentum ${\hat N}$ and the spin angular momentum ${\hat S}$ \cite{gamma}, ${\hat E}$ and ${\hat B}$ are the electric and magnetic fields, ${\hat d}$ is the electric dipole moment of the molecule and $\mu_B$ is the Bohr magneton. If the quantization axis is oriented along the magnetic field direction, the electric-field-induced interaction is represented as \cite{Schmelcher}
\beq\label{e3}
\hat{E}\cdot\hat{d} = Ed\frac{4\pi}{3}\sum_q Y^{\star}_{1q}({\bm r})
Y_{1q}({\bm E}),
\eeq
where ${\bm r}$ and ${\bm E}$ define the directions of the interatomic axis and the electric field with respect to the magnetic field axis.
The Hamiltonian ({\ref{e2}}) depends on the angle between the electric and magnetic fields through the $Y_{1q}({\bm E})$ spherical harmonics.
We use the most accurate potential for He--CaH interactions \cite{pes} and 
the dipole moment  $d=2.94$ D \cite{dipmom}. 

In the absence of the spin-rotation interaction and in collinear fields, the CaH molecule is characterized by the projections of
${\hat N}$ and ${\hat S}$ -- denoted by $M_N$ and $M_S$ -- on the magnetic field axis. The spin-rotation interaction couples
states with different $M_N$ and $M_S$ while
the interaction with electric fields couples rotational states of opposite parities.
We will label molecular states by the quantum numbers $N$ and $M_S$ that describe the
leading contribution in the corresponding eigenvector of the Hamiltonian (\ref{e2}). 
If the magnetic and electric fields are parallel, the total angular momentum projection $M=M_N+M_S+M_\ell$, where $M_\ell$ is the projection of the rotational 
angular momentum ${\hat \ell}$ of the He--CaH complex, is conserved. The scattering calculations can be carried out in a cycle over $M$. If the magnetic and electric fields are not parallel, the interaction with electric fields couples states with different $M_N$ and the total angular momentum projection $M$ is not
conserved. The dimension of the scattering problem is thus greatly increased. 
We include in our calculations the molecular states with $N\le 7$ and $\ell \le 8$. In the absence of external fields, this would correspond to a series of 15 calculations with less than 128 coupled states. If the magnetic and electric fields are parallel, the number of coupled equations to solve is  808 for $M = 1/2$. The same calculation in non-parallel fields would involve integration of 10368 coupled differential equations.
As this is beyond our computational resources, we had to reduce the basis to $N\le 5$ and $\ell \le 6$ and integrate
3528 equations to study  the effect of crossed fields.
In this Letter, we consider collisions of molecules initially in the lowest-energy low-magnetic-field seeking state $|N=0,M_S=1/2\rangle$. 
Collisional spin relaxation leads to inelastic $|N=0,M_S=1/2\rangle \to |N=0,M_S=-1/2\rangle$ transitions that, as we show, can be controlled by both magnetic and electric fields.

The upper panel of Fig. 1 shows the variation of the spin relaxation cross sections with an electric field at a magnetic field strength $B$ = 0.1 T and collision energy $\epsilon = 0.5$ K. We present calculations for the CaH molecule with the rotational constant $B_e = 6.05$~K, CaD with $B_e = 3.11$~K, and a hypothetic molecule with $B_e=1$~K and $\gamma = 5.9 \times 10^{-3}$~cm$^{-1}$.  Increasing the electric field suppresses the spin relaxation.
We can explain this effect as follows. Our previous calculations \cite{krems,krems2} demonstrated that the spin relaxation
in rotationally ground-state $^2\Sigma$ molecules occurs through coupling to rotationally excited molecular levels and is
determined by the spin-rotation interaction in the rotationally excited molecular states.
The couplings between the rotational energy levels are smaller in 
molecules with larger rotational constants. 
The energy separation between the magnetic levels of the ground 
$N = 0$ and first excited $N = 1$ states increases with the electric field 
strength \cite{bretislav}. Applying an electric field is thus equivalent to increasing the 
rotational constant of the molecule. 
In addition, the electric 
field splits the states with different values of $M_N$ in the rotationally 
excited levels, which decreases the effective spin-rotation interaction. 
Both of these factors reduce the effective couplings between the spin 
states of the ground rotational state, which leads to suppression
of the spin relaxation, as shown in Fig. 1. 


Fig. 1 demonstrates two important observations: (1) Electric fields have a dramatic effect on collisional spin relaxation in multiple partial wave scattering, i.e. at {\it cold} temperatures easily achievable in cryogenic experiments; (2) the suppression  of the spin relaxation is stronger in molecules with smaller rotational constants. This result indicates that sympathetic or evaporative cooling of heavy molecules in magnetic traps may be possible in the presence of strong electric fields. 

Spin-changing transitions in $^2\Sigma$ molecule - atom collisions are nearly forbidden at zero temperature in weak magnetic fields. The presence of electric fields induces direct couplings between the Zeeman states and enhances the cross sections for the spin relaxation to a great extent (Fig.~2). 
Cross sections for inelastic transitions at ultralow energies must be inversely proportional to the collision velocity $v$ -- the relation known as the Wigner law. Fig.~2 shows that the magnitude of the electric field determines the onset of the ultracold $1/v$ dependence of the cross sections. The upturn of the cross section computed at zero electric field occurs at $10^{-10}$ cm$^{-1}$ and is outside the scale of the graph.

The Zeeman levels corresponding to different rotational states cross at high magnetic fields. For CaD, these crossings occur at magnetic fields of about 4.5 T. In the presence of electric fields these crossings transform into avoided crossings and molecular properties such as orientation and alignment
are sensitive to the magnitudes of the fields near the avoided crossings \cite{bretislav,H2}. The upper panel of Fig.~3 shows the cross sections for the spin
relaxation as functions of the magnetic field at different electric fields near the crossings. At zero electric field, the cross section is insensitive to the crossings and
varies slowly with the magnetic field. In the presence of an electric field, however, the cross sections show sharp peaks corresponding to the positions of the
avoided crossings that mix the spin-up and spin-down Zeeman states. The electric field dependence of the cross sections at fixed magnetic fields (lower panel
of Fig.~3) shows similar resonances. 

Rotating the electric field breaks the axial symmetry of the problem and induces couplings between states with different total angular momentum projections.
While the perturbations due to rotating electric fields are small, they may alter the collision dynamics near avoided crossings.
Figure 4 presents the cross section for the spin relaxation at electric and magnetic fields near an avoided crossing as a function of the angle between the fields. The Hamiltonian matrix changes and the spin relaxation cross sections vary significantly. Fig.~4 thus demonstrates that inelastic molecular collisions can be manipulated not only by varying the strength of applied electric and magnetic fields, but also by changing the relative orientation of the superimposed fields.

In summary, we have shown that inelastic spin relaxation in collisions involving $^2\Sigma$ molecules can be enhanced or suppressed by suitably chosen combinations of superimposed electric and magnetic fields. 
The mechanisms of electric field control of molecular spin dynamics are based on an intricate interplay of intramolecular spin-rotation couplings and molecule - field interactions. We have demonstrated that inelastic collisions of $^2\Sigma$ molecules with atoms can be controlled not only in the limit of ultracold single partial wave scattering, but also at warmer temperatures achievable in cryogenic cooling experiments. Electric fields can modify the range of ultracold Wigner scattering by inducing couplings between states otherwise uncoupled. 
We have shown that cross sections for inelastic Zeeman transitions can be manipulated by varying not only the strength, but also the relative orientation of the electric and magnetic fields.

Our results bear significant implications for research in, at least, three different areas of physics and chemistry. Fig.~1 indicates that applying strong electric fields in the EDM measurement experiments will make heavy $^2\Sigma$ molecules more stable against collisional spin decoherence. Using electric fields to induce spin flipping may also find applications in quantum computation schemes, where communications between quantum bits are enabled by spin-dependent interactions. Cooling diatomic molecules to ultracold temperatures in a magnetic trap is believed to be very challenging due to rapid spin relaxation in molecule - molecule collisions. If weak magnetic dipole interaction is ignored, the mechanism of the spin relaxation in collisions of rotationally ground-state $^2\Sigma$ molecules is the same as in atom - molecule collisions.
This follows from the equations in Sec. IIC of Ref. [14]. If either the spin-rotation interaction in
Eq. (20) or the couplings between the different rotational states in Eqs. (25) 
and (26) are omitted, the different spin states of the $N=0$ level are 
completely uncoupled and the spin relaxation in rotationally ground-state 
molecules cannot occur.
The different rotational states are coupled in Eq. (24) by the anisotropy 
of the molecule - molecule interaction potential. The interaction potential operator is independent of the electric 
field. Electric fields mix the different rotational states, which may result in a change of the coupling matrix elements
between the field-dressed rotational states. This change of the already 
large atom - molecule and molecule - molecule interaction anisotropy should, however, not be significant. 
Our results thus indicate that electric fields may be used to facilitate not only the sympathetic cooling of molecules in atomic buffer gases, but also the evaporative cooling of dilute molecular ensembles in magnetic traps by inhibiting collisional spin relaxation. 

Finally, our calculations suggest that spin-forbidden chemical reactions may be amenable to electric field control at low temperatures. Consider, for example, a chemical reaction between an atom in the $^2S$ electronic state and a $^2\Sigma$ molecule. If the atom and molecule are both confined in a magnetic trap, they are initially in the state with the total spin $S=1$. The interactions between $^2S$ atoms and $^2\Sigma$ molecules in the maximally stretched spin state are typically characterized by large reaction barriers. The interactions in the singlet spin state are often
strongly attractive, leading to short-range minima and insertion reactions \cite{avoird}.
As demonstrated here, electric fields can induce or suppress the spin 
reorientation in molecules, which may induce or suppress the singlet-triplet non-adiabatic transitions in 
pre-reactive complexes and result in enhancement of chemical reactivity at low temperatures.


We acknowledge stimulating discussions with Jonathan Weinstein. 
This work is supported by the Natural Sciences and Engineering Research Council (NSERC) of Canada and the Killam postdoctoral research fellowship.

\begin{figure}[ht]
\label{f1}
\begin{center}
\includegraphics[width=5.9cm]{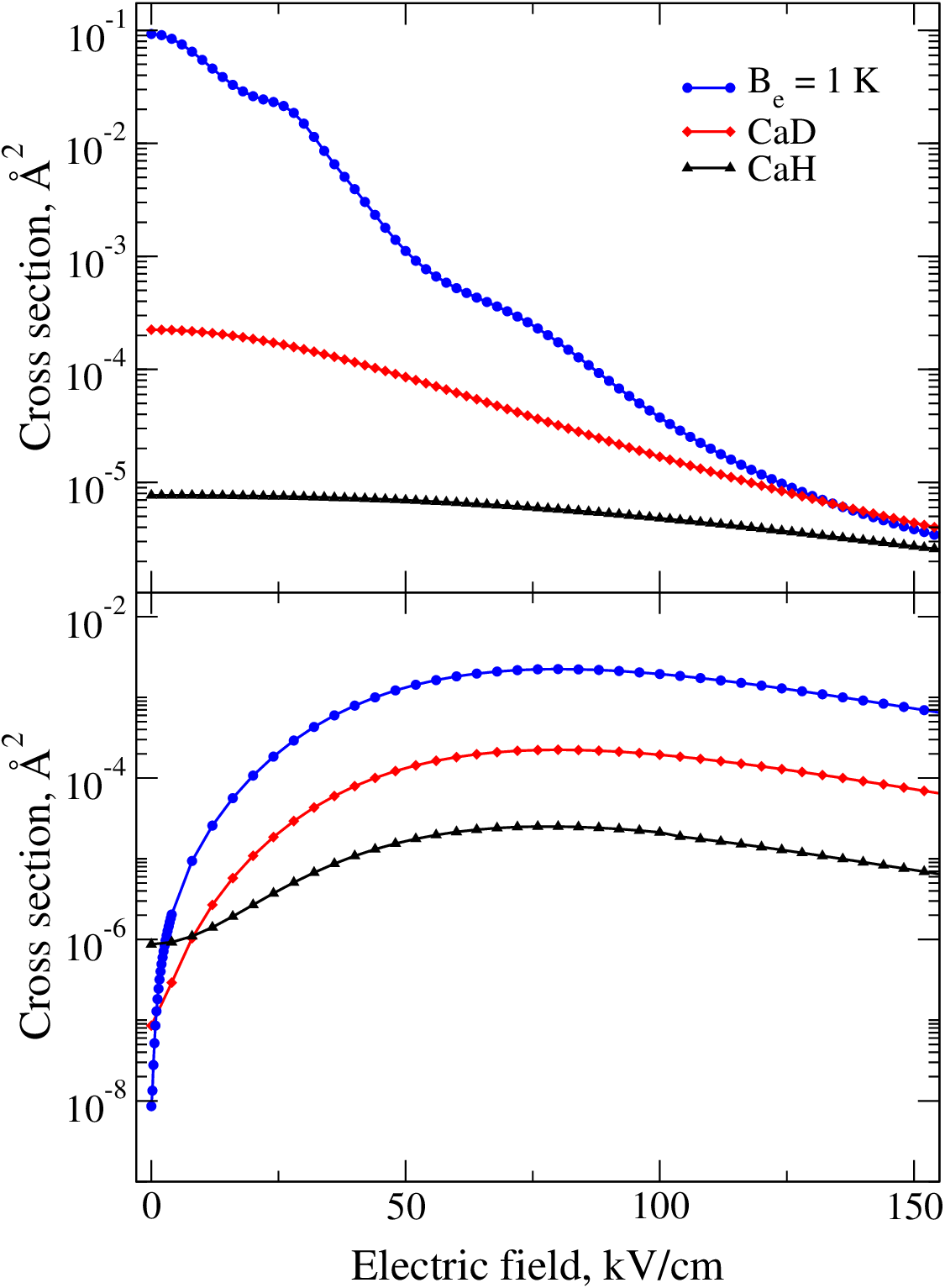}
\end{center}
\caption{Electric field dependence of spin relaxation cross sections 
in molecule - He atom collisions  at a magnetic field of $0.1$ T. The magnetic and electric fields are parallel.  
Upper panel: triangles --  the rotational constant  of the molecule $B_e = 6.05$ K (CaH); 
diamonds -- $B_e=3.11$ K (CaD); squares - a hypothetical molecule with $B_e=1$ K; the collision energy is 0.5 K. 
Lower panel: CaD--He scattering at ultracold collision energies of $10^{-3}$ K (triangles), $10^{-5}$ K (diamonds) and $10^{-7}$ K (circles).
}
\end{figure}

\begin{figure}[ht]\label{f1}
\includegraphics[width=5.9cm]{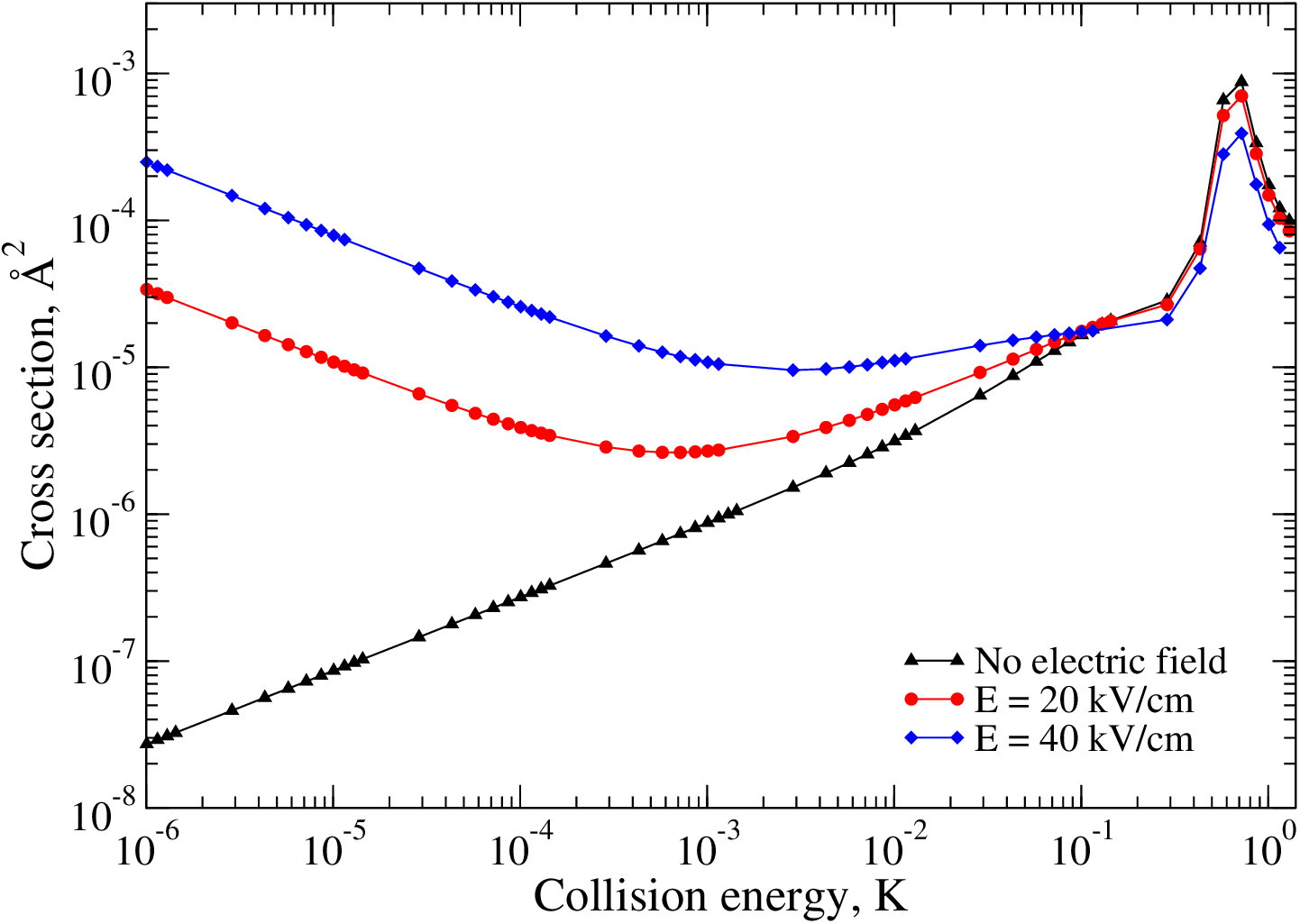}
\caption{Collision energy dependence of spin relaxation cross sections in CaD--He collisions at a magnetic field
of  $ 0.1$ T and zero electric field (triangles), $E = 20$ kV/cm (circles), and $E=40$ kV/cm
(diamonds). The magnetic and electric fields are parallel. 
}
\end{figure}

\begin{figure}[ht]
\label{f1}
\begin{center}
\includegraphics[width=5.9cm]{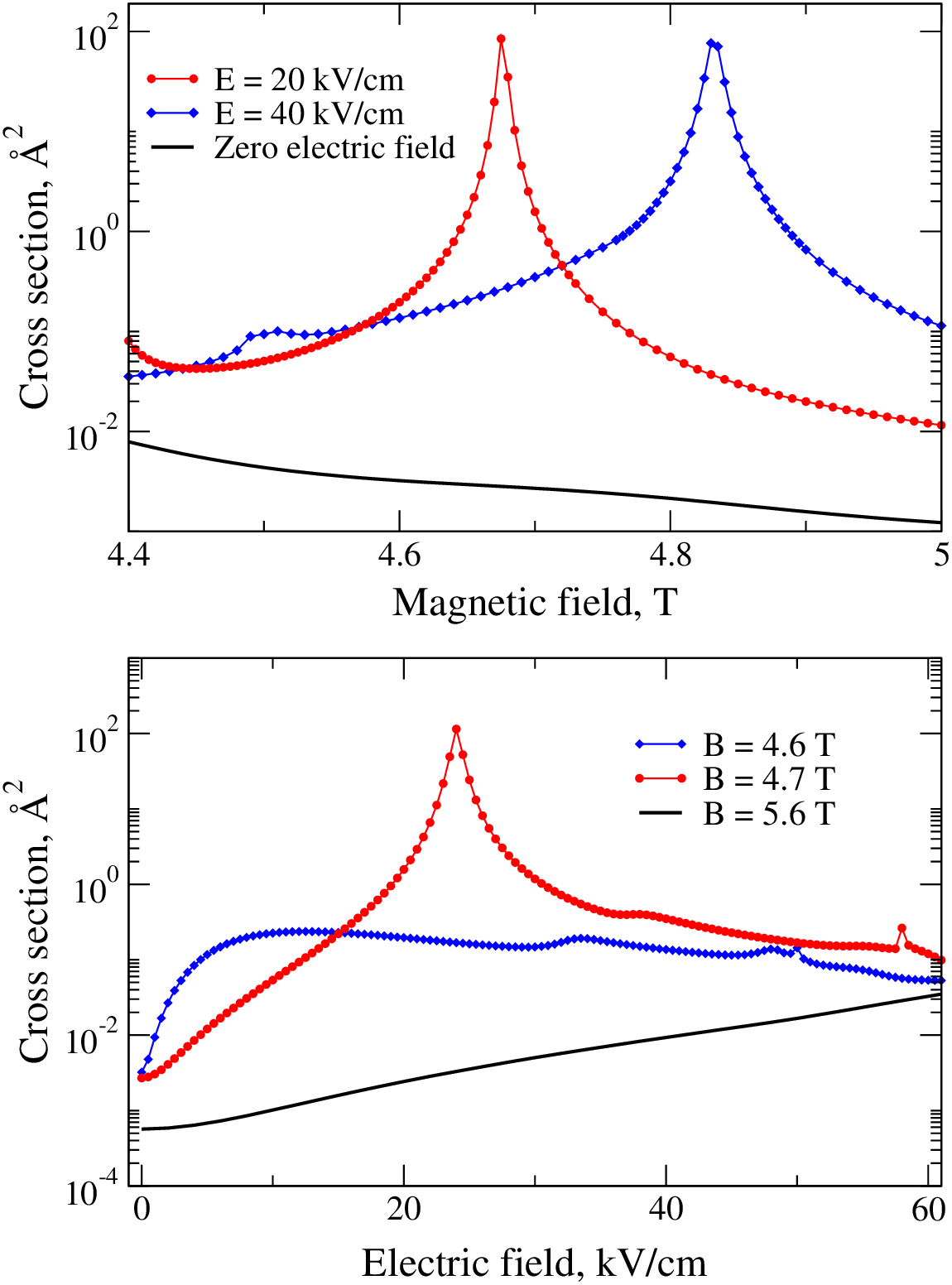}
\end{center}
\caption{
Upper panel:  Magnetic field dependence of spin relaxation cross sections in CaD--He collisions 
at fixed electric fields of $0$ (full line), $20$ (circles), and $40$ kV/cm (diamonds) near an avoided crossing of the
$N=0$ and $N=1$ spin sublevels.
Lower panel: Electric field dependence of spin relaxation cross sections at magnetic fields of
$4.6$ T (diamonds), $4.7$ T (circles), and $5.6$ T (full line). The magnetic and electric fields
are parallel.}
\end{figure}

\begin{figure}[ht]\label{f1}
\includegraphics[width = 0.30\textwidth]{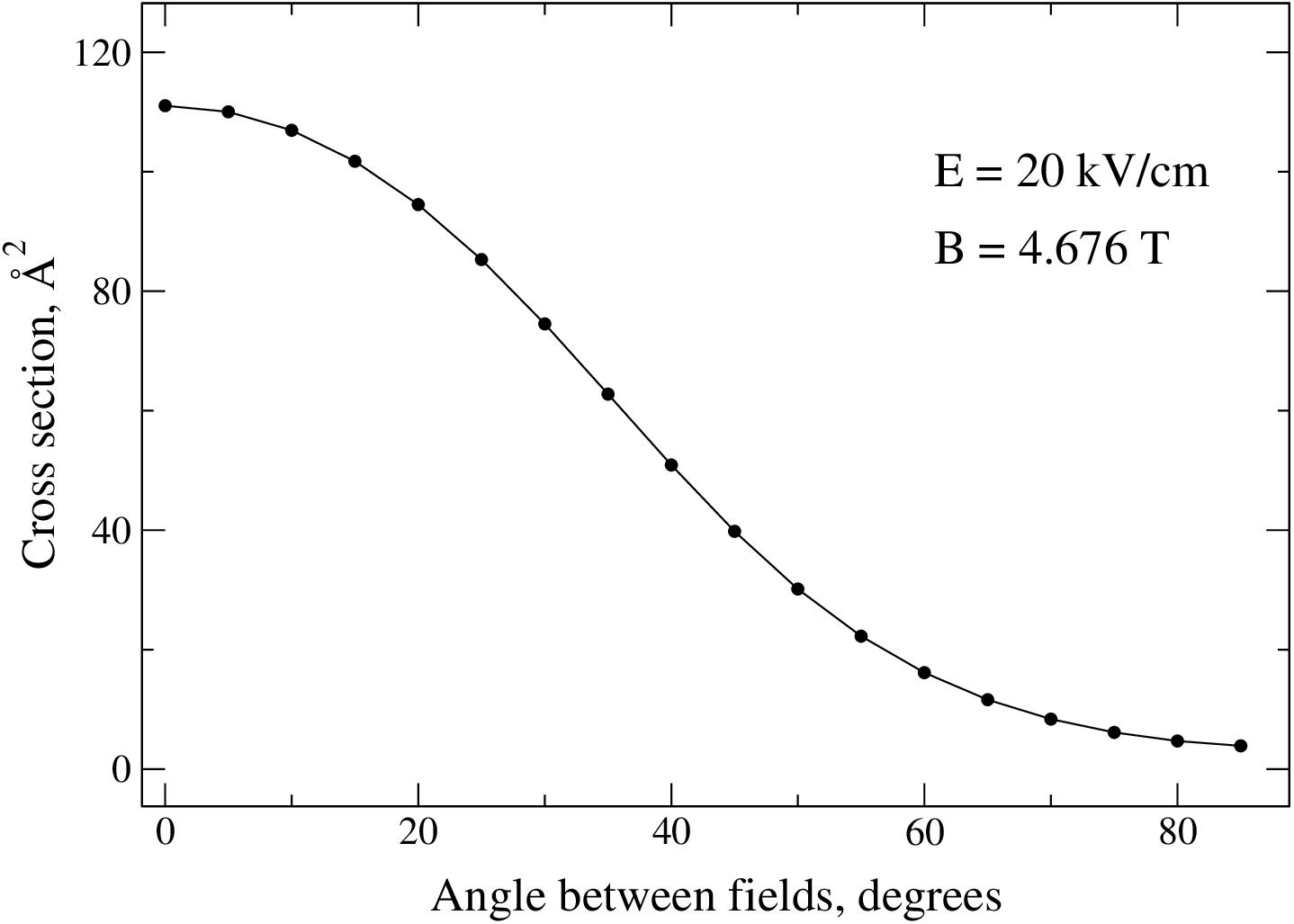}
\caption{Spin relaxation cross sections in CaD--He collisions at $B=4.676$ T
and $E=20$ kV/cm versus the angle between the electric and magnetic fields. The
collision energy is 0.5 K.}
\end{figure}


\begin{thebibliography}{99}

\bibitem{zoller}
P. Zoller and J. I. Cirac, 
Physics Today, {\bf 38} (2004). 

\bibitem{hinds}
E. Hinds, Phys. Scr. {\bf T70}, 34 (1997);
J. J. Hudson \etal,
Phys. Rev. Lett. {\bf 89}, 023003 (2002).

\bibitem{moshe}
M. Shapiro, and P. Brumer,  2003,  {\it ``Principles of Quantum Control of Molecular Processes''} (John Wiley and Sons, Inc., New Jersey). 

\bibitem{demille}
D. DeMille, \prl{88}{067901}{2002}.

\bibitem{mitia}
R. Barnett \etal,
\prl{96}{190401}{2006}.

\bibitem{zoller2}
A. Micheli, G. K. Brennen, and P. Zoller, 
Nature Physics {\bf 2}, 341 (2006).   

\bibitem{demille2}
D. DeMille \etal,
Phys. Rev. A {\bf 61}, 052507 (2000).

\bibitem{john}
D. Egorov \etal,
Phys. Rev. A {\bf 63}, 030501 (2001)

\bibitem{demille3}
M. G. Kozlov and D. DeMille,
Phys. Rev. Lett. {\bf 89}, 133001 (2002). 

\bibitem{irpc}
R. V. Krems,  Int. Rev. Phys. Chem. {\bf 24}, 99 (2005) -- and references therein.  

\bibitem{bohn_e}
A. V. Avdeenkov and J. L. Bohn,
Phys. Rev. Lett. {\bf 90}, 043006 (2003);
C. Ticknor and J. L. Bohn,
{\bf 72}, 032717 (2005); 
A. V. Avdeenkov, M. Kajita, and J. L. Bohn,
Phys. Rev. A {\bf 73}, 022707 (2006).

\bibitem{bohn_m}
A. Volpi and J. L. Bohn,
Phys. Rev. A {\bf 65}, 052712 (2002).

\bibitem{krems}
R. V. Krems \etal,
Phys. Rev. A {\bf 68}, 051401(R) (2003); 
R. V. Krems, \prl{93}{013201}{2004};
H. Cybulski \etal, \jcp{122}{094307}{2005}. 


\bibitem{krems2}
R.V. Krems and A. Dalgarno,
\jcp{120}{2296}{2004}.


\bibitem{buffer-gas-loading}
J. M. Doyle \etal,
Phys. Rev. A, {\bf 52},  R2515 (1995).

\bibitem{bretislav}
B. Friedrich and D. Herschbach, 
Phys. Chem. Chem. Phys. {\bf 2}, 419 (2000).

\bibitem{roman2}
R.V. Krems \etal,
\pra{67}{060703(R)}{2003}.

\bibitem{mizushima}
K.-P. Huber and G. Herzberg, {\it Constants of Diatomic Molecules} (Van Nostrand Reinhold,
New York, 1979).

\bibitem{gamma}
The value of $\gamma$ used is $0.0415$ cm$^{-1}$ for
CaH and $0.021$ cm$^{-1}$ for CaD \cite{mizushima}

\bibitem{Schmelcher}
R. Gonz{\'a}lez-F{\'e}rez and P. Schmelcher, \pra{69}{023402}{2004};
\pra{71}{033416}{2005}

\bibitem{pes}
G. C. Groenenboom, N. Balakrishnan, \jcp{118}{7380}{2003};
N. Balakrishnan \etal,
\jcp{118}{7386}{2003}.

\bibitem{dipmom}
T. C. Steimle, J. Chen, and J. Gengler, \jcp{121}{829}{2004}.



\bibitem{H2}
B. Friedrich, H.-G. Rubahn, and N. Sathyamurthy, \prl{69}{2487}{1992}.

\bibitem{avoird}
P. Sold{\'a}n and J. M. Hutson, \prl{92}{163202}{2004}.

\end{thebibliography}
\end{document}